\documentclass[11pt]{article}

\usepackage{authblk}
\usepackage{fullpage}
\usepackage[utf8]{inputenc}
\usepackage[T1]{fontenc}
\usepackage{microtype}
\usepackage[dvipsnames]{xcolor}
\usepackage[subpreambles]{standalone}
\usepackage{tikz}
\usetikzlibrary{patterns}

\usepackage{hyperref}
\hypersetup{
    colorlinks=true,
    linkcolor=olive,
    citecolor=CadetBlue
}

\usepackage[backend=biber,natbib=true,citestyle=numeric-comp]{biblatex}
\bibliography{citations.bib}

\usepackage[light, largesmallcaps, easyscsl, nowarning]{kpfonts}
\usepackage{amsthm,amssymb,mathabx}
\newtheorem{theorem}{Theorem}
\newtheorem{conjecture}[theorem]{Conjecture}
\newtheorem{lemma}[theorem]{Lemma}
\theoremstyle{definition}
\newtheorem{definition}[theorem]{Definition}
\newtheorem{observation}[theorem]{Observation}
\newtheorem{corollary}[theorem]{Corollary}

\newcommand{\MMM}{\textscsl{mmm}}
\newcommand{\EDS}{\textscsl{eds}}
\newcommand{\UGC}{\textsc{ugc}}

\newcommand{\MBB}{\textscsl{mbb}}

\usepackage{import}

\title{Tight Approximation Ratio for Minimum Maximal Matching}
\date{}
\author{Szymon Dudycz}
\author{Mateusz Lewandowski}
\author{Jan Marcinkowski}
\affil{Institute of Computer Science, University of Wrocław \\
    \texttt{\{firstname.lastname\}@cs.uni.wroc.pl}}

\begin{document}
    \maketitle

    \begin{abstract}
        We study a combinatorial problem called \emph{Minimum Maximal Matching},
        where we are asked to find in a general graph the smallest matching
        that can not be extended. We show that this problem is hard to
        approximate with a constant smaller than 2, assuming the Unique Games
        Conjecture.

        As a corollary we show, that Minimum Maximal Matching
        in bipartite graphs is hard to approximate with constant
        smaller than \(\frac{4}{3}\), with the same assumption. With a stronger
        variant of the Unique Games Conjecture --- that is Small Set Expansion
        Hypothesis --- we are able to improve the hardness result up to the
        factor of \(\frac{3}{2}\).
    \end{abstract}


\section{Introduction}

    Matchings are one of the most central combinatorial structures in theory of
    algorithms. A routine computing them is a basic puzzle used in numerous
    results in Computer Science (like Christofides algorithm). Various variants
    of matchings are studied extensively. Their computation complexity status is
    usually well-known and some techniques discovered when studying matchings
    are afterwards employed in other problems.

    As we know since 1961, all natural variants of perfect matchings and maximum matchings
    can be found in polynomial time, even in general graphs. Here we study a
    different problem --- \emph{Minimum Maximal Matching} (\MMM{}). The task is
    --- given graph \(G\), to find an inclusion-wise maximal matching \(M\) with
    the smallest cardinality (or weight in the weighted version).

    \subsection{Related Work}
    The \MMM{} problem was studied as early as 1980, when Yannakakis and Gavril
    showed, that it is NP-hard even in some restricted cases~\cite{YG80}. Their
    paper also presents an equivalence of \MMM{} and \emph{Minimum Edge
    Dominating Set} (\EDS{}) problem, where the goal is to find minimum
    cardinality subset of edges \(F\), such that every edge in the graph is
    adjacent to some edge in \(F\). Every maximal matching is already an edge
    dominating set, and any edge dominating set can be easily transformed to a
    maximal matching of no larger size. This equivalence does not hold for the
    weighted variants of the problem.

    It is a well known, simple combinatorial fact, that one maximal matching in
    any graph can not be more than twice as large as another maximal matching.
    This immediately gives a trivial 2-approximation algorithm for \MMM{}.
    Coming up with 2-approximation in the weighted variant of either of the
    problems is more challenging. In 2003, Carr, Fujito, Konjevod and Parekh
    presented a \(2\frac{1}{10}\)-approximation algorithm for a weighted \EDS{}
    problem~\cite{CarrFKP00}. Later the approximation was improved to 2 by
    Fujito and Nagamochi~\cite{FujitoN02}.

    Some algorithms aiming at approximation ratio better then 2 were also
    developed for the unweighted problem. Gotthilf, Lewenstein and Rainschmidt came
    up with a \(2-c\frac{\log n}{n}\)-approximation for the general
    case~\cite{GotthilfLR08}. Schmied and Viehmann have a better-than-two
    constant ratio for dense graphs~\cite{SchmiedV12}.

    Finally, hardness results need to be mentioned. In 2006 Chleb\'{\i}k and
    Chleb\'{\i}kov\'{a} proved, that it is NP-hard to approximate the problem
    within factor better than~\(\frac{7}{6}\)~\cite{ChlebikC06}. The result was
    later improved to \(1.18\) by Escoffier, Monnot, Paschos, and
    Xiao~\cite{EscoffierMPX15}. \(\frac{3}{2}\)-hardness results depending on
    \UGC{} were also obtained~\cite{SchmiedV12,EscoffierMPX15}.

    \subsection{Unique Games Conjecture}
    Unique Games Conjecture, since being formulated by Khot in
    2002~\cite{Khot02a}, has been used to prove hardness of approximation of
    many problems. For the survey on \UGC{} results see~\cite{Khot10}.

    Many hardness results obtained from Unique Games Conjecture match previously
    known algorithms, as is the case, for example, of \emph{Vertex Cover},
    \emph{Max Cut} or \emph{Maximum Acyclic Subgraph}. Therefore, it is
    appealing to use it to obtain new results. While \UGC{} is still open,
    recently a related 2--2-Games Conjecture has been proved~\cite{KhotMS18}, in
    consequence proving Unique Games Conjecture with partial completeness. This
    result provides some evidence towards validity of Unique Games Conjecture.

    Basing on Unique Games Conjecture we are able to prove the main result of
    our paper.

    \begin{theorem}
        Assuming Unique Games Conjecture, it is NP-hard to approximate
        Minimum Maximal Matching with constant better than 2.
    \end{theorem}

    The proof of this theorem relies on the UGC-hardness proof for
    Vertex Cover of Khot and Regev~\cite{kr07}. In essence, we endeavour
    to build a matching over the vertices of Vertex Cover.

    As a side-effect of our proof, hardness of approximating \emph{Total Vertex
    Cover} follows. In this problem the goal is to find a subset \(W\) of
    vertices, which is a Vertex Cover and every vertex in \(W\) is incident to
    at least one other vertex in \(W\).

    \begin{corollary}
        Assuming Unique Games Conjecture, it is NP-hard to approximate
        Total Vertex Cover with constant better than 2.
    \end{corollary}

    The Minimum Maximal Matching problem does not seem to be easier on bipartite
    graphs. All the algorithms mentioned above are defined for general graphs and
    we are not aware of any ways to leverage the bipartition of the input graph.
    At the same time, our hardness proof only works for general graphs. With some
    observations we are able to achieve a hardness result for bipartite
    graphs, which, however, is not tight.

    \begin{theorem}
        Assuming Unique Games Conjecture, it is NP-hard to approximate
        bipartite Minimum Maximal Matching with constant better than
        \(\frac{4}{3}\).
    \end{theorem}

    \subsection{Obtaining a Stronger Result}
    The studies on Unique Games Conjecture and hardness of approximation of
    different problems have led to formulating different hypotheses
    strengthening upon \UGC{} --- among them the \emph{Small Set Expansion
    Hypothesis} proposed by Raghavendra and Steurer~\cite{RaghavendraS10}, and
    another conjecture --- whose name is not yet established and so far the name
    \emph{Strong UGC} is used --- formulated by Bansal and
    Khot~\cite{BansalK09}. A competent discussion on differences between the two
    conjectures can be found in~\cite[Appendix~C]{Manurangsi18}.

    To improve our result on bipartite graphs, we construct a reduction from a
    problem called \emph{Maximum Balanced Biclique} (\MBB{}), where --- given a
    bipartite graph --- the goal is to find a maximum clique with the same number
    of vertices on each side of the graph. Hardness of approximation results
    suitable for our reduction have been found starting from both the Small Set
    Expansion Hypothesis~\cite{Manurangsi18} and Strong
    UGC~\cite{BhangaleGHKK17}.

    \begin{theorem}
        Assuming Small Set Expansion Hypothesis (or Strong Unique Games
        Conjecture), it is NP-hard to approximate Bipartite Minimum Maximal
        Matching with a constant better than \(\frac{3}{2}\).
    \end{theorem}


\section{Revisiting the Khot-Regev Reduction}
In their paper~\cite{kr07} Khot and Regev prove the \(\UGC \)-hardness of
approximating Minimum Vertex Cover within a factor smaller than \(2\). In
this section we look at parts of their proof more closely.

Their reduction starts off with an alternative formulation of \(\UGC
\)\footnote{In their paper, Khot and Regev call this formulation ``Strong Unique
Games Conjecture''. Since then, however, the same name has been used to refer
another formulation, as in~\cite{BansalK09}, we decided to minimise confusion by
not recalling this name.}, which, they show, is a consequence of the standard
variant.

\subsection{Khot-Regev Formulation of Unique Games Conjecture}
This formulation talks about a variant of Unique Label Cover problem described
variables, \(E\) are the edges and \(\Psi_{x_1,x_2}\) defines a constraint
by a tuple \(\Phi = \left(X, R, \Psi, E\right)\). \(X\) is a set of
for every pair of variables connected by an edge. A constraint is a
permutation \(\Psi_{x_1,x_2} \in R \leftrightarrow R\) meaning that if
\(x_1\) is labelled with a colour \(r \in R\), \(x_2\) must be labelled with
\(\Psi_{x_1,x_2}(r)\).

A \emph{\(t\)-labelling} is an assignment of subsets \(L(x)\) of size
\(|L(x)| = t\) to the variables. A constraint \(\Psi_{x_1,x_2}\) is
satisfied by the t-labelling \(L\) if there exists a colour \(r \in L(x_1)\)
such that \(\Psi_{x_1,x_2}(r) \in L(x_2)\).

\begin{conjecture}[Unique Games Conjecture]\label{conj:strong-ugc}

    For any \(\xi, \gamma > 0 \) and \( t \in \mathbb{N}\) there exists some
    \(|R|\) such that it is NP-hard to distinguish, given an instance \(\Phi =
    \left(X, R, \Psi, E\right)\) which category it falls into:
    \begin{itemize}
        \item (\textsc{yes} instance): There exists a labelling
        (\emph{1-labelling}) \(L\) and a set \(X_0 \subseteq X\), \(|X_0|
        \geqslant (1 - \xi)|X|\), such that \(L\) satisfies all constraints
        between vertices of \(X_0\).
        \item (\textsc{no} instance): For any t-labelling \(L\) and any set
        \(X_0 \subseteq X\), \(|X_0| \geqslant \gamma|X|\), not all
        constraints between variables of \(X_0\) are satisfied by \(L\).
    \end{itemize}
\end{conjecture}

\subsection{Weighted Vertex Cover}
The next step is a reduction from the \(\UGC \) to the Minimum Vertex Cover
problem. Given an instance \(\Phi = \left(X, R, \Psi, E\right)\) 
 of Unique Label Cover problem, as described above, we build a graph
\(G_{\Phi}\).

For every variable in \(x \in X\) we create a \emph{cloud} \(\mathscr{C}_x\)
of \(2^{|R|}\) vertices. Each vertex corresponds to a subset of labels and
is denoted by \(\left(x, S\right) \in |X| \times \mathcal{P}(R)\). The
weight of a new vertex \((x, S)\) is equal to
\[
    \mu(|S|) = \frac{1}{|X|} \cdot p^{|S|}{(1-p)}^{|R \setminus S|}
\]
where \(p = \frac{1}{2}-\varepsilon \) (there is a bias towards smaller
sets). The total weight of \(G_{\Phi}\) is thus equal to~\(1\).

Next, we connect the vertices \((x_1, S_1)\) and \((x_2, S_2)\) if the
labellings \(S_1\) and \(S_2\) do not satisfy the constraint
\(\Psi_{x_1,x_2}\). Two lemmas are proved.

\begin{lemma}[{\cite[][Sec.~4.2]{kr07}}]\label{lem:kr07-yes}
    If \(\Phi \) was a \textsc{yes} instance, the graph \(G_{\Phi}\) has an
    independent set of weight at least \(\frac{1}{2} - 2\varepsilon \).
\end{lemma}
\begin{proof}
    The instance \(\Phi \), being a \textsc{yes} instance, has a labelling
    \(L\) assigning one colour \(r_x\) to each variable \(x\). We know, that
    there is a large set \(X_0\) of variables (\(|X_0| \geqslant \left(1 -
    \xi\right)|X|\)), such that all constraints between variables of \(X_0\)
    are satisfied by \(L\).

    We now define
    \[
        \mathcal{IS} = \big \{ (x, S)\ \big|\ x \in X_0,\, r_x \in S \big \}
    \]
    and claim, that \(\mathcal{IS}\) is an independent set in \(G_{\Phi}\).
    For any two variables \(x_1\) and \(x_2\) of \(X_0\) we know, that
    \[
        \Psi_{x_1,x_2}(r_{x_1}) = r_{x_2}.
    \]
    Indeed, if we then take the sets of labels \(S_1 \ni r_1\) and \(S_2 \ni
    r_2\), they do satisfy the constraint for the variables \(x_1\),
    \(x_2\). Hence, there is no edge between \((x_1, S_1)\) and \((x_2,
    S_2)\).

    Finally, the weight of \(\mathcal{IS}\) is equal to
    \[
        \begin{split}
            w\left(\mathcal{IS}\right) & =
            \sum_{x \in X_0} \left(\sum_{S \subseteq R, S \ni r_x} w(x, S) \right) = 
            \sum_{x \in X_0} \left(\frac{1}{|X|} \cdot \sum_{k=1}^{|R|} \binom{|R|-1}{k-1} \cdot p^k \cdot {\left(1-p\right)}^{|R|-k} \right) \\ & =
            \sum_{x \in X_0} \left(p \cdot \frac{1}{|X|} \cdot \sum_{k=0}^{|R|-1} \binom{|R|-1 }{k} \cdot p^k \cdot {\left(1-p\right)}^{|R|-1-k} \right) \\ & =
            \sum_{x \in X_0} \left(p \cdot \frac{1}{|X|} \cdot {\left(p + \left(1-p\right)\right)}^{|R|-1}\right) \\ & =
            \frac{|X_0|}{|X|} \cdot p \geqslant (1-\xi)(\frac{1}{2}-\varepsilon) >
            \frac{1}{2} - 2\varepsilon.
        \end{split}
    \]
\end{proof}

The most of their paper is dedicated to proving the following key lemma.
\begin{lemma}[{\cite[][Sec.~4.3]{kr07}}]\label{lem:kr07-no}
    If \(\Phi \) is a \textsc{no} instance, it does not have an independent
    set of weight larger than~\(2\gamma \).
\end{lemma}

Since the Minimum Vertex Cover is a complement of the Maximum
Independent Set, we see that it is hard to distinguish between graphs with
Minimum Vertex Cover of the weight \(\frac{1}{2} + 2\varepsilon \) and those,
where Minimum Vertex Cover weights \(1 - 2\gamma \).

\subsection{Notation}
Throughout this paper we are going to use \(\Phi \) as an instance of
Unique Label Cover problem that we are translating to \(G_{\Phi}\). The weight
function \(w\) on vertices and bias function \(\mu \) is going to be recalled,
as well as the constants \(\varepsilon \) and \(\gamma \). When \(\Phi \) is a
\textsc{yes} instance, we are going to refer to the set \(X_0\) as in
Conjecture~\ref{conj:strong-ugc}, and use the independent set \(\mathcal{IS}\)
from Lemma~\ref{lem:kr07-yes}.


\section{Weighted Minimum Maximal Matching}\label{sec:weightedmmm}
Let us now modify their reduction. The graph \({G}^{\prime}_{\Phi}\) gets
additional edges between vertices \((x, S_1)\), \((x, S_2)\) if \(S_1 \cap
S_2 = \varemptyset \) --- they do not assign the same colour to the variable
\(x\). Clearly, the Lemmas~\ref{lem:kr07-yes} and~\ref{lem:kr07-no} still
hold for \({G}^{\prime}_{\Phi}\).

Moreover, we introduce the weight function on the edges.
\[
    w_{+}\left(\left(x_1, S_1\right), \left(x_2, S_2\right)\right)
    \overset{\text{def}}{=\joinrel=}
    w\left(x_1, S_1\right) + w\left(x_2, S_2\right)
\]

We will now show the similar statements are true for the Minimum Maximal
Matching as for the independent set.
\begin{lemma}\label{lem:wei-yes}
    If \(\Phi \) was a \textsc{yes} instance, the Minimum Maximal Matching
    in \(\left({G}^{\prime}_\Phi, w_+\right)\) weights at most~\(\frac{1}{2}
    + 2\varepsilon \).
\end{lemma}
\begin{lemma}\label{lem:wei-no}
    If \(\Phi \) was a \textsc{no} instance, the Minimum
    Maximal Matching in \(\left({G}^{\prime}_{\Phi}, w_+\right)\) weights at
    least~\(1 - 2\gamma \).
\end{lemma}

These lemmas altogether will give us the theorem.

\begin{theorem}
    Assuming the Unique Games Conjecture, for any \( \epsilon > 0 \) it is
    NP-hard to distinguish between graphs with Maximal Matching of weight
    \(\frac{1}{2} + \epsilon \) and those, where every Maximal Matching weights
    at least \( 1 - \epsilon \).
\end{theorem}

This in turn means, that --- assuming UGC --- a polynomial-time approximation algorithm
with a factor better than \(2\) can not be constructed.

\begin{proof}[Proof of Lemma~\ref{lem:wei-yes}]

    Let us construct a matching \(M\) in \(G^{\prime}_{\Phi}\). The matching
    will only consist of the edges between vertices corresponding to the same
    variable in \({\Phi}\). First we define the part of \(M\) restricted to
    \(X_0\).
    \[
        M_0 = \Big \{{(x, S_1) \sim (x, S_2)\ \big|\ x \in X_0 \,\wedge \,
        S_1 \uplus S_2 = R\setminus \{r_x\}}\Big \}
    \]
    For vertices in clouds corresponding to variables outside of \(X_0\) we
    define
    \[
        M_1 = \Big \{{(x, S_1) \sim (x, S_2)\ \big|\ x \in X_0 \,\wedge \,
        S_1 \uplus S_2 = R}\Big \}
    \]
    The matching \(M\) will be the union of \(M_0 \) and \(M_1 \).

    We can observe, that the vertices matched by \(M\) are exactly those,
    that do not belong to \(\mathcal{IS}\). Hence,
    \[
        \begin{split}
            w_+(M) & \leqslant w(G^{\prime}_{\Phi}) - w(\mathcal{IS})
            \leqslant 1 - \left( \frac{1}{2} - 2\varepsilon \right) =
            \frac{1}{2} + 2\varepsilon
        \end{split}
    \]

    Moreover, since the vertices of \(M\) compose a vertex cover, \(M\) is a
   maximal matching.
\end{proof}

\begin{proof}[Proof of Lemma~\ref{lem:wei-no}]

    Let \(M\) be any maximal matching. The vertices matched by \(M\),
    \(V(M)\) form a vertex cover. Hence, the weight of \(M\) is going to be
    at least as large as the weight of the Minimum Vertex Cover. From
    Lemma~\ref{lem:kr07-no} we know, that if \({\Phi}\) was a \textsc{no}
    instance, \(G^{\prime}_{\Phi}\)'s Minimum Vertex Cover weights at least
    \(1 - 2\gamma \).
\end{proof}

\section[Towards the Unweighted MMM: Fractional Matchings]{Towards the Unweighted MMM:\@ Fractional Matchings}\label{sec:frac-matching}
    A natural way to reduce a weighted variant of a problem to the unweighted
    would often be to assume that the weights are integral (that can be achieved
    by rounding them first at a negligible cost) and copying every vertex as
    many times, as its weight would suggest. This simple strategy will not
    however work with instances from previous section, where we were matching
    pairs of vertices of different weights. Such a matching does not easily
    translate to the graph with vertex copies. In order to extend our
    approximation hardness proof to Minimum Maximal Matching problem in
    unweighted graphs, we thus need first to modify our weighted reduction a
    bit. The structure remains the same, but the weight of each edge is now
    defined to be the minimum of the weights of its endpoints.
    \[
        w_{\min}\left(\left(x_1, S_1\right), \left(x_2, S_2\right)\right)
        \overset{\text{def}}{=\joinrel=}
        \min{ \{w(x_1, S_1), w(x_2, S_2) \}}
    \]

    Similarly to the reasoning presented in the previous section, when
    \(G^{\prime}_{\Phi}\) is a \textsc{yes} instance, we will want to construct
    a matching and argue that it is maximal using a known vertex cover.

    \begin{definition}
        A \emph{fractional matching} is an assignment of values to variables
        \(x_e\) corresponding to edges, such that for every edge \(e\) \(x_e
        \leqslant w_{\min}(e)\) and for every vertex \(v\), the sum
        \(\sum_{(v,w) \in E} x_{(v, w)} \leqslant w(v)\).
    \end{definition}

    \begin{definition}
        A fractional matching \emph{saturates} the vertex \(v\) if \(\sum_{(v,w)
        \in E} x_{(v, w)} = w(v)\).
    \end{definition}

    As we know already, when \(\Phi \) is a \textsc{yes} instance, there is
    a vertex cover in \(G^{\prime}_{\Phi}\) composed of all vertices except those
    in \(\mathcal{IS}\).

    \begin{lemma}\label{lem:fra-mat}
        If \(\Phi \) was a \textsc{yes} instance, a fractional matching exists
        that leaves all vertices in \(\mathcal{IS}\) unmatched and saturates all
        the other vertices.
    \end{lemma}

    \subsection{Proving Lemma~\ref{lem:fra-mat}}
    Our matching will again only match vertices in the same clouds. Let us first
    concentrate on vertices in the cloud \(\mathscr{C}_x\) corresponding to a
    variable \(x \not\in X_0\). The matching needs to saturate every vertex in
    \(\mathscr{C}_x\).

    The fractional matching \(F\) can be viewed as a real-valued vector and
    will be a sum of three matchings. The first one is defined similarly to
    \(M_1\) in Lemma~\ref{lem:wei-yes}.

    \[
        F^0\big((x, S_1), (x, S_2)\big) =
        \begin{cases}
            w_{\min}\left((x, S_1), (x, S_2)\right), &
                \mbox{if } S_1 \uplus S_2 = R \\
            0, & \mbox{otherwise}
        \end{cases}
    \]

    Recalling, that the weight function \(w\), defined on vertices, has a bias
    towards smaller sets, we can state the following.

    \begin{observation}
        \(F^0\) saturates all vertices \((x, S)\in \mathscr{C}_x\) such that
        \(|S| \geqslant \frac{|R|}{2}\).
    \end{observation}

    Let us now pick \(0 < k < \frac{|R|}{2}\) and look at the layer
    \(\mathscr{C}_x^k = {\big \{ (x, S) \ \big | \ |S| = k \big \}}\). The graph
    is symmetric, and \(F^0\) saturates every vertex by the same amount ---
    \(\mu(|R|-k) = \frac{1}{|X|}  p^{|R|-k} {(1-p)}^k\). In order to build a matching
    \(F^1\), that saturates all vertices in the layer we build a bipartite
    graph \(\mathcal{B}^k\) out of \(\mathscr{C}_x^k\)\footnote{A significantly
    more crude approach is possible, that just uses every edge equally.}.

    \begin{definition}
        For every set \(S\) of size \(k\), \(\mathcal{B}^k\) has two vertices,
        \(S^L\) and \(S^R\). \(S_1^L\) is connected with \(S_2^R\) if \(S_1 \cap
        S_2 = \varemptyset \).
    \end{definition}
    The graph \(\mathcal{B}_k\) is in fact a \emph{Bipartite Kneser Graph}. As
    proved in~\cite{Mutze2017}, it has a Hamiltonian cycle \(\mathcal{H}_k\). We
    are using this cycle to define \(F^1\) --- for every edge connecting the
    sets \(S_1\) and \(S_2\) in \(\mathcal{H}_k\) we lay the weight of
    \[
        F^1\left((x, S_1), (x, S_2)\right) =
            \frac{1}{4} \left( \mu(k) - \mu(|R|-k) \right)
    \]
    on the edge connecting them in \(\mathscr{C}_x^k\).

    To saturate the vertices \((x, \varemptyset)\) (for \(x \not \in X_0\)), we
    must realize that all these vertices form a clique in which we can find a
    Hamiltonian Cycle \(\mathcal{H}_\emptyset \). Let us define \(F^2\)
    \[
        F^2\big((x_1, \varemptyset), (x_2, \varemptyset)\big) =
        \begin{cases}
            \frac{\mu(0) - \mu(|R|)}{2}, &
                \mbox{for } \{x_1, x_2\} \in \mathcal{H}_\emptyset \\
            0, & \mbox{otherwise}
        \end{cases}
    \]

    \begin{lemma}
        \(F^0 + F^1 + F^2\) saturates all vertices in \(\mathscr{C}_x^k\).
    \end{lemma}
    \begin{proof}
        We look at the vertex \((x, S)\). For \(0 < |S| < \frac{|R|}{2} \), the
        Hamiltonian Cycle \(\mathcal{H}_k\) visits every set exactly twice (once
        \(S^L\) and once \(S^R\)), using four edges incident to it. Hence, the
        total contribution of \(F^0\) and \(F^1\) is equal to
        \[
            \mu(|R|-k) +
            4 \cdot \frac{1}{4} \left(\mu(k)-\mu(|R|-k) \right) =
            \mu(k) = w(x, S).
        \]
        \(F^0\) contributes \(\mu(|R|)\) to the vertex \((x, \varemptyset)\),
        while \(F^2\) contributes \(2 \cdot \frac{\mu(0) - \mu(|R|)}{2}\), hence
        that vertex is also saturated.

        Finally, vertices with \(S = \varemptyset \) are saturated by \(F^0 +
        F^2\).
    \end{proof}

    \subsubsection[When x in X0]{When \(x\in X_0\)}
    We proceed similarly as for vertices not in \(X_0\).
    For the cloud \(\mathscr{C}_x\) when \(x \in X_0\), our first matching
    \(F^0\) is taking the labeling of the variable \(x\) into account.
    Similarly to Lemma~\ref{lem:wei-yes}, we match \((x, S_1)\) and \((x, S_2)\)
    if \(S_1 \uplus S_2 = R \setminus \{r_x\} \), thus saturating the larger of
    the sets.

    Again, the layer \(\mathscr{C}_x^k\) for \(k < \frac{|R|-1}{2}\), composed
    of sets not containing \(r_x\), is a Bipartite Kneser Graph, and we use its
    Hamiltonian cycle to define \(F^1\).

    Also the vertices \((x, \varemptyset)\) for \(x \in X_0\) form a clique.
    Once again, we can use the Hamiltonian Cycle in that clique to define
    \(F^2\).

\section{Unweighted MMM}\label{sec:cardinality}
    Starting with a graph \(G_{\Phi}\) with the weight function \(w\) on the
    vertices, and any precision parameter \(\rho > 0\), we are going to
    construct an unweighted graph \(G_{\Phi}^{\rho} = \left(V^{\rho},
    E^{\rho}\right)\). The resulting graph size is polynomial in \(|\Phi|\) and
    \(\frac{1}{\rho}\).

    \begin{definition}
        Let \(n = |V(G_{\Phi})| \cdot \frac{1}{\rho}\). For every \(v \in
        V(G_{\Phi})\) we set \(n_v = \lceil n \cdot w(v) \rfloor \). The new set
        of vertices is going to consist of multiple copies of original vertices;
        for each vertex \(v\), we add \(4 \cdot n_v\) copies.
        \[
            V^\rho = \big \{ \left< v, i \right> \ \big |\  v \in V(G_\Phi),
                i \in \{1, \dots, 4\cdot n_v\} \big \}.
        \]
        The edges are going to connect each pair of copies of vertices connected
        in \(G_\Phi \).
        \begin{gather*}
            E^\rho = \big \{ \{ \left<v_1, i_1\right>, \left<v_2, i_2 \right> \}
            \ \big |\ {\{v_1, v_2\}} \in E(G_\Phi), \\
                i_1 \in [4\cdot n_{v_1}],
                i_2 \in [4\cdot n_{v_2}] \big \}.
        \end{gather*}
    \end{definition}

    This construction has been presented in~\cite{Dinur02}. It is shown, that
    any vertex cover \(C \subset G_{\Phi}\) yields a \emph{product vertex cover}
    \(C^{\rho} = \bigcup_{v \in C} \{v\}\times [4\cdot n_v] \). Moreover, every
    minimal vertex cover in \(G_{\Phi}^{\rho}\) is a product vertex
    cover~\cite[][Proposition~8.1]{Dinur02}.

    As before, we are now going to prove two lemmas witnessing the completeness
    and soundness of our reduction.

    \begin{lemma}[Soundness]\label{lem:card-soundness}
        If \(\Phi \) was a \textsc{no} instance, for every maximal matching
        \(M\) in \(G_{\Phi}^{\rho}\)
        \[
            2\cdot |M| > |V(G_{\Phi}^{\rho})| \left( 1 - 2\gamma - \rho \right).
        \]
    \end{lemma}
    \begin{proof}
        Take any maximal matching \(M\). The \(2\cdot|M|\) vertices matched by
        it form a vertex cover \(C\). Let \(C_-\) be a minimal vertex cover
        obtained by removing unneeded vertices from \(C\). As presented
        in~\cite{Dinur02}, \(C_-\) is a product vertex cover, which means, there
        is a vertex cover \(C_w\) in \(G_{\Phi}\) with weight
        \[
            w(C_w) < \frac{|C_-|}{V(G_{\Phi}^{\rho})} + \rho \leqslant
            \frac{|C|}{V(G_{\Phi}^{\rho})} + \rho.
        \]
        On the other hand, from Lemma~\ref{lem:kr07-no} we have, that \(w(C_w) >
        1 - 2\gamma \).
    \end{proof}

    \begin{figure}
        \centering{}
        \begin{tikzpicture}[
        every node/.style={draw,pattern=crosshatch dots,circle},
        every label/.append style={rectangle, font=\Large}]
    \tikzstyle{layer}=[lightgray,thick,fill=gray!30,rounded corners=20pt]
    \tikzstyle{dim}=[draw=black!50,pattern color=black!50];

    \clip (-10,-6) rectangle (10,6);
    \draw[lightgray,thick,fill=lightgray!20] (0,0) ellipse (10 and 12);

    \draw[layer] (-8.5,3) rectangle (8.5, 5);
    \draw[layer] (-9,.5) rectangle (9, 2.5);
    \draw[layer] (-8.7,-2) rectangle (8.7, 0);
    \draw[layer] (-8,-4.5) rectangle (8, -2.5);

    \node[inner sep=6pt,label={below:\((x, S)\)}] (xs) at (2,-3.5) {};
    \node[label={above:\((x, R\setminus(S \cup \{r_x\}))\)}] (xcs) at (0, 4)
    {};

    \draw[line width=4pt] (xs) to[out=80,in=-30]
    node[draw=none,fill=none,midway,label={right:\(F^0\)}] {} (xcs);

    \node[dim,inner sep=6pt] (xs1) at (-6.5,-3.5) {};
    \node[dim,inner sep=6pt] (xs2) at (-4,-3.5) {};
    \node[dim,inner sep=6pt] (xs3) at (-1.5,-3.5) {};
    \node[dim,inner sep=6pt] (xs4) at (5.5,-3.5) {};

    \draw[thick] (xs) to[out=220,in=320] (xs1);
    \draw[thick] (xs) to[out=130,in=60]
        node[draw=none,fill=none,midway,label={above:\(F^1\)}] {}
        (xs2);
    \draw[thick] (xs) to[out=160,in=20] (xs3);
    \draw[thick] (xs)
    to[out=20,in=150] (xs4);

\end{tikzpicture}
        \caption{A close-up look at the resulting matching in a cloud of
            vertices corresponding to the variable \(x \in X_0\). The fractional
            matchings \(F^0\) and \(F^1\) constructed in
            Section~\ref{sec:frac-matching} can be discretised to match all the
            copies of respective vertices. }\label{fig:card-match}
    \end{figure}
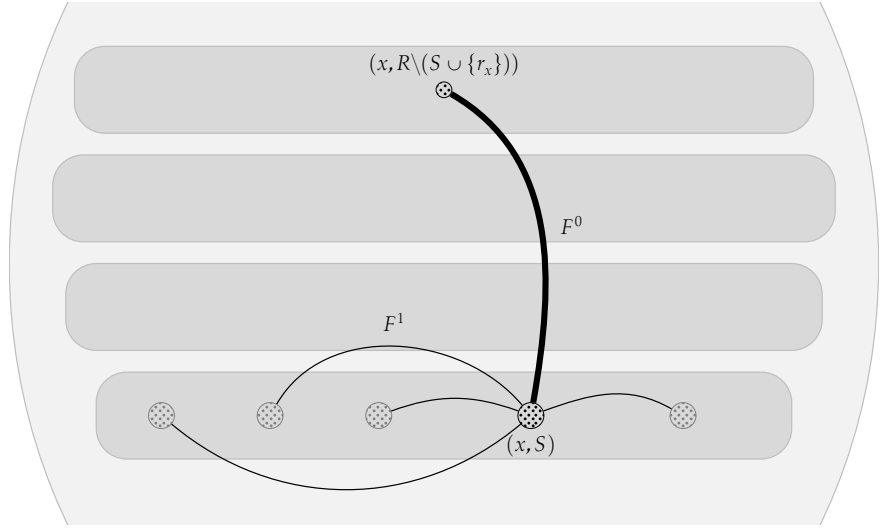
    \begin{lemma}[Completeness]\label{lem:card-completeness}
        If \(\Phi \) was a \textsc{yes} instance, a maximal matching \(M\)
        exists in \(G_{\Phi}^{\rho}\) with
        \[
            2\cdot|M| < |V(G_{\Phi}^{\rho})| \left( \frac{1}{2} + 2\varepsilon +
                \rho \right).
        \]
    \end{lemma}
    \begin{proof}
        Take \(F\), a fractional matching on \((G_{\Phi}, w_{\min})\)
        constructed in Lemma~\ref{lem:fra-mat}. When \(F^0\) matches vertices
        \(u = (x, S_1)\) and \( v = (x, S_2)\) with some weight \(F^0(u, v)\),
        we are going to match \(4 \cdot \lceil F^0(u, v) \rfloor \) copies of
        \(u\) and \(v\) using parallel edges.

        Let us focus on a vertex \(u = (x, S)\not\in\mathcal{IS}\) belonging to
        a vertex cover of \(G_{\Phi}\), with \(0 < |S| < \frac{|R|}{2}\). It is
        matched by \(F^0\) to \((x, S')\), which leaves \(4\left(\lceil w(x, S)
        \rfloor - \lceil w(x, S') \rfloor \right) \) vertices in
        \(G_{\Phi}^{\rho}\) unmatched. This number is divisible by 4, which
        allows us to match all the copies of vertices in the Bipartite Kneser
        Graph according to \(F^1\) (see Fig~\ref{fig:card-match}).

        Finally, the number of unmatched copies of the \((x, \varemptyset)\)
        vertices is divisible by 2. We can thus replicate \(F^2\) to match all
        the remaining copies of these vertices.

        Since we are matching every node in a vertex cover of the graph
        \(G_{\Phi}^{\rho}\), our matching is maximal and its cardinality is half
        of the cardinality of the vertex cover.

        \[
            |M| = \frac{1}{2} \left(V(G_{\Phi}^{\rho}) - |IS|^{\rho}\right) <
            \frac{1}{2} V(G_{\Phi}^{\rho}) \left(1 -
                \left(\frac{1}{2} - 2\varepsilon - \rho\right)\right)
        \]
    \end{proof}

    \subsection{Hardness of Total Vertex Cover}
    In Lemmas~\ref{lem:card-soundness} and~\ref{lem:card-completeness} we proved
    \UGC{}-hardness of the following problem. For any \(\epsilon>0\), given a
    graph \(G\) with \(n\) vertices it is hard to distinguish if:
    \begin{itemize}
        \item (\textsc{yes} instance) \(G\) has a Maximal Matching of size
            smaller than \(n \left( \frac{1}{4} + \epsilon \right) \).
        \item (\textsc{no} instance) \(G\) has no Vertex Cover of size
            smaller than \(n \left( 1 - \epsilon \right)\).
    \end{itemize}

    Vertices matched in \MMM{} form a Total Vertex Cover, so in the \textsc{yes}
    case there is a Total Vertex Cover of size smaller than \(n \left(
    \frac{1}{2} + \epsilon \right) \). On the other hand, every Total Vertex
    Cover is a Vertex Cover, so in the \textsc{no} case there is no Total Vertex
    Cover of size smaller than \(n \left( 1 - \epsilon \right)\).

    Therefore, Total Vertex Cover is \UGC{}-hard to approximate with constant
    better than 2.


\section{Hardness of Bipartite MMM}\label{sec:bipartisation}
    In this section we will perform a natural reduction to prove the following
    theorem.

    \begin{theorem}\label{thm:bipartite43}
        Assuming the Unique Games Conjecture, for any \(\epsilon > 0 \) it is
        NP-hard to distinguish between balanced bipartite graphs of \(2n\)
        vertices:
        \begin{itemize}
            \item (\textsc{yes} instance) with a Maximal Matching of size
                smaller than \(n \left( \frac{1}{2} + \epsilon \right) \).
            \item (\textsc{no} instance) with no Maximal Matching of size
                smaller than \(n \left( \frac{3}{2} - \epsilon \right)\).
        \end{itemize}
    \end{theorem}

    We will start with the graph \(G_{\Phi}^{\rho}\) defined in
    Section~\ref{sec:cardinality}. The bipartite graph \(H_{\Phi}\) has two
    copies \(v^{l}\) and \(v^{r}\) of every vertex \(v \in G_\Phi^\rho \). The
    vertices \(u^{l}\) and \(v^r\) are connected with an edge if there is an
    edge \((u, v)\) in \(G_\Phi^\rho \). \(n\) is going to be equal to
    \(|V(G_\Phi^\rho)|\). We will call this construction \emph{bipartisation} of
    an undirected graph.

    It is easy to see, that if \(\Phi \) is a \textsc{yes} instance of the
    Unique Label Cover problem, we can use the matching from
    Lemma~\ref{lem:card-completeness} (\(M\) in \(G_\Phi^\rho \)) to produce
    a maximal matching in \(H_\Phi \). For every edge \((u, v)\in M\) we will
    put its two copies, \((u^l, v^r)\) and \((v^l, u^r)\) into the matching. The
    resulting matching size is thus equal to \(2 \cdot |M| < n (\frac{1}{2} +
    \epsilon) \).

    \subsection{Covering with Paths}

    In order to analyse the \textsc{no} case, we need to look at the bipartite
    instance and its matchings from another angle. For any matching in
    \(H_{\Phi}\), we will view its edges as directed edges in \(G_\Phi^{\rho}\)
    --- the vertices on the left will be viewed as \emph{out} vertices, and
    those on the right as \emph{in} vertices. The graph \(G_{\Phi}^{\rho}\) will
    thus be covered with directed edges. Every vertex will be incident to at
    most one outgoing and one incoming edge, which means that the edges will
    form a structure of directed paths and cycles. The set of these paths and
    cycles will be called \(\mathscr{P}(M)\) for a matching \(M\).

    \begin{observation}\label{obs:bipartisation-paths}
        If \(M\) is a maximal matching, every path \(P \in \mathscr{P}(M)\) has
        a length \(|P| \geqslant 2\).
    \end{observation}
    \begin{proof}
        Assume, that for a maximal matching \(M\) in \(H_\Phi \) there is a
        length-one path \(P = {(u, v)} \in \mathscr{P}(M)\). This means, that
        the vertices \(v^l\) and \(u^r\) are unmatched in \(M\) --- yet, they
        are connected with an edge, that can be added to the matching (that
        would form a length-2 cycle in \(\mathscr{P}(M)\)).
    \end{proof}

    We will now use this observation to prove the relation between maximal
    matchings in \(H_\Phi \) and vertex covers in \(G_\Phi^\rho \).

    \begin{lemma}
        For any maximal matching \(M\) in \(H_\Phi \), there exists a vertex
        cover \(C\) in \(G_\Phi^{\rho}\) of size
        \(|C| \leqslant \frac{3}{2}|M|\).
    \end{lemma}
    \begin{proof}
        We will construct the vertex cover using paths and cycles of
        \(\mathscr{P}(M)\). For every \(P \in \mathscr{P}(M)\) we add all the
        vertices of \(P\) into \(C\). When \(P\) is a cycle, it contains as many
        vertices as edges. A path has at most \(\frac{3}{2}\) as many vertices
        as edges, since its length is at least 2.
    \end{proof}

    As shown in Lemma~\ref{lem:card-soundness}, when \(\Phi \) is a \textsc{no}
    instance, the Minimum Vertex Cover in \(G_\Phi^\rho \) has at least \(n (1 -
    \epsilon)\) vertices. The Minimum Maximal Matching in \(H_{\Phi}\) must in
    this case have at least \(\frac{2}{3}n(1 - \epsilon) > n(\frac{2}{3} -
    \epsilon)\) edges.

    The hardness coming from Theorem~\ref{thm:bipartite43} is, that assuming
    UGC, no polynomial-time algorithm will provide approximation for Minimum
    Maximal Matching with a factor \(\frac{4}{3} - \epsilon \) for any
    \(\epsilon > 0\).

\section{Stronger Result for Bipartite Graphs}

We are able to obtain a stronger hardness of approximation result for bipartite
graphs, but it assumes a slightly stronger conjecture. In this Section we 
will show how to prove hardness assuming Small Set Expansion 
Hypothesis~\cite{RaghavendraS10}, but the same result can be obtained from
Strong Unique Games Conjecture --- it requires replacing Lemma~\ref{lem:manurangsi}
with a corresponding hardness of \MBB{} from~\cite{BhangaleGHKK17}. In order to describe the reasoning,
let us now characterise Minimum Maximal Matching solutions in bipartite graphs
and focus on the conjecture later.

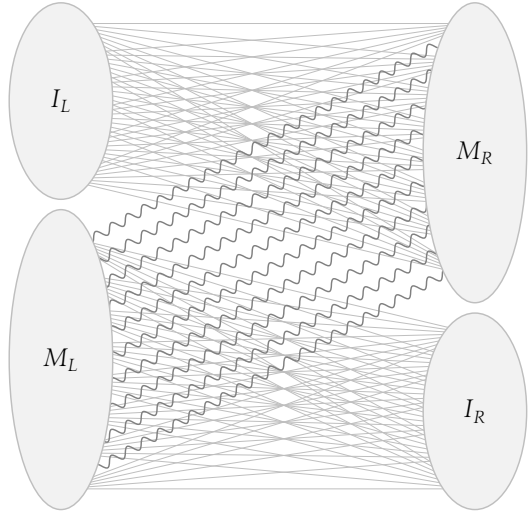
\begin{figure}
	\centering{}
        \begin{tikzpicture}[
            every label/.append style={rectangle, font=\Large}]

        \clip (-6,-6) rectangle (6,6);

        \foreach \i in {0,...,5} {
            \pgfmathsetmacro{\l}{4.5 - \i * .6};
            \foreach \j in {0,...,7} {
                \pgfmathsetmacro{\r}{4.5 - \j * .7};
                \draw[lightgray] (-4, \l) -- (4, \r);
                \draw[lightgray] (4, -\l) -- (-4, -\r);
            }
        }

        \foreach \i in {0,...,9} {
            \pgfmathsetmacro{\l}{-4.5 + \i * 5/10};
            \pgfmathsetmacro{\r}{0 + \i * 5/10};
            \draw[gray,thick, decorate, decoration=snake] (-4, \l) -- (4, \r);
        }

        \draw[lightgray,thick,fill=lightgray!20] (-4, 3) ellipse (1 and 1.9);
        \draw[lightgray,thick,fill=lightgray!20] (-4, -2) ellipse (1 and 2.9);

        \draw[lightgray,thick,fill=lightgray!20] (4, 2) ellipse (1 and 2.9);
        \draw[lightgray,thick,fill=lightgray!20] (4, -3) ellipse (1 and 1.9);

        \node[draw=none] at (-4, 3) {\Large{\(I_L\)}};
        \node[draw=none] at (4, -3) {\Large{\(I_R\)}};
        \node[draw=none] at (-4, -2) {\Large{\(M_L\)}};
        \node[draw=none] at (4, 2) {\Large{\(M_R\)}};

    \end{tikzpicture}
    \caption{A bipartite graph with Minimum Maximal Matching. The (optimal)
        matching is drawn with stronger, snaky lines. The unmatched vertices
        form a balanced bipartite anti-biclique in the
        graph.}\label{fig:biclique}
\end{figure}

Imagine, our balanced bipartite graph \(H\) has a perfect matching. Clearly, its
Minimum Maximal Matching \(M\) has at least \(\frac{n}{2}\) edges (where \(n\)
is a number of vertices on either side). \(M\) allows us to divide vertices of
\(H\) into the sets \(M_L, M_R\) (matched in \(M\)) and \(I_L, I_R\)
(unmatched). In previous sections we used the fact, that the set \(I_L \cup
I_R\) is an independent set to bound the size of Minimum Maximal Matching in the
\textsc{no} case. Here, we notice that \(I_L, I_R\) is a balanced anti-biclique
--- its complement is a bipartite clique. Clearly, if a bipartite graph of
\(2n\) vertices has no \(\overline{K}_{\delta n,\delta n}\) anti-biclique, its
Minimum Maximal Matching must be larger than \((1-\delta)n\).

We recall a recent result by Manurangsi, who has proved the following lemma.
\begin{lemma}[{\cite[][Lemma 2]{Manurangsi18}}]\label{lem:manurangsi}
	Assuming the \emph{Small Set Expansion Hypothesis}, for every \(\epsilon >
	0\) it is NP-hard to distinguish, given a bipartite graph \(G=(A\cup B, E)\)
	with \(|A|=|B|\), which category it falls into:

	\begin{itemize}
		\item (\textsc{yes} case) \(\exists K_A \subset A, K_B \subset B\) such
			that \(E\restriction_{K_A \cup K_B} =
			K_{|V(G)|(\frac{1}{2}-\epsilon), |V(G)|(\frac{1}{2}-\epsilon)}\).
			Namely, there is a balanced biclique in \(G\) using almost half of
			vertices.
		\item (\textsc{no} case) \(\forall K_A \subset A, K_B \subset B\,
			|K_A|=|K_B| > \epsilon |A| \implies \exists a\in K_A, b\in K_B\
			(a,b)\not\in E(G)\). Namely, there is no balanced biclique using
			more than \(\epsilon \)-fraction of vertices.
	\end{itemize}
\end{lemma}

We will now use this lemma to prove \(\frac{3}{2}\) approximation hardness for
Minimum Maximal Matching problem in bipartite graphs. Take an instance \(G\)
from Lemma~\ref{lem:manurangsi}. Let \(n\) be the number of vertices on one
side. Our modified graph \(G' = ((A \cup A') \cup (B \cup B'), \bar{E} \cup
E')\) is created by adding \(n(\frac{1}{2}+\epsilon)\) vertices each to both
sides, so \(|A'| = |B'| = (\frac{1}{2}+\epsilon) n\). To produce the set of
edges we first take complement of \(E\) on \(A\) and \(B\). This way bicliques
become anti-bicliques. Next we add \(E' = A' \times B \cup A \times B' \cup A'
\times B'\) (we connect the new vertices with every vertex from \(G\) and with
each other).

\begin{lemma}\label{lem:bip-sseh-yes}
	If \(G\) has a \(K_{n(\frac{1}{2}-\epsilon),n(\frac{1}{2}-\epsilon)}\)
	biclique \(K_A \times K_B\), there is a maximal matching with \(n(1 +
	2\epsilon)\) edges in \(G'\).
\end{lemma}
\begin{proof}
	We can match vertices of \(A\setminus K_A\) with \(B'\) and \(A'\) with
	\(B\setminus K_B\). All the remaining vertices, \(K_A \cup K_B\), form an
	anti-biclique in \(G'\) so the matching is maximal.
\end{proof}

\begin{lemma}\label{lem:bip-sseh-no}
	If \(G\) has no biclique \(K_{\epsilon n, \epsilon n}\), every maximal
	matching in \(G'\) contains at least \(n (\frac{3}{2} - \epsilon)\)
	edges.
\end{lemma}
\begin{proof}
	It suffices to argue that \(G'\) does not have a large anti-biclique. Since
	all vertices in \(A'\) are connected with everyone, only one of them can
	belong to the anti-biclique. The same applies to \(B'\). The remainder of
	the anti-biclique would form a biclique in \(G\). The largest anti-biclique
	in \(G'\) can therefore have \(\epsilon n\) vertices on each side.
\end{proof}

\section{Conclusion}
	We would like to finish by discussing potentially interesting open problems.
	Natural question following our result on \MMM{} 
	is whether other hardness results for
	Vertex Cover also hold for \MMM{}. In particular, it is known that Vertex
	Cover on \(k\)-hypergraphs is hard to approximate with a constant better
	than \(k\)~\cite{kr07}. Also, the best known NP-hardness of Vertex Cover is
	\(\sqrt{2}\), following the reduction from 2--2 Games
	Conjecture~\cite{KhotMS17}, which has been recently proven~\cite{KhotMS18}.

	Both of these reductions are very similar to Khot and Regev's
	\UGC{}-hardness of Vertex Cover. As such they can be used to prove
	corresponding hardnesses of weighted \MMM{}, by following similar approach
	as in Section~\ref{sec:weightedmmm}. They differ, however, in the
	choice of the weight function of vertices, which turns out to be crucial in
	terms of unweighted \MMM{}. These weight functions have bias towards
	bigger sets, so construction described in Section~\ref{sec:frac-matching}
	can not be used for these problems.

	As such, the best known NP-hardness of \MMM{} remains \(1.18\) 
	by Escoffier, Monnot, Paschos, and Xiao~\cite{EscoffierMPX15} and 
	it is an open problem, whether it can be improved using 2--2 Games
	Conjecture.

	In case of bipartite \MMM{}, there remains a gap between our \(\frac{3}{2}\)-hardness
	and best known constant approximation algorithm, which has ratio 2. Showing 
	that bipartite \MMM{} is hard to approximate with a constant better than 2
	would immediately imply tight hardness of 
	\emph{Maximum Stable Matching with Ties}~\cite{HuangIMY15}. On the other hand,
	there are no results for \MMM{} leveraging restriction to bipartite graphs.
	Thus, a potential better than 2 approximation algorithm for bipartite graphs
	would be interesting for showing structural difference between \MMM{} in
	bipartite and general graphs.



    \printbibliography{}

\end{document}